# Catalytic upgrading of hydrothermal liquefaction biocrudes: Different challenges for different feedstocks

Daniele Castello*, Muhammad Salman Haider, Lasse Aistrup Rosendahl

Department of Energy Technology, Aalborg University, Pontoppidanstræde 111, 9220 Aalborg Øst,

Denmark

*Corresponding author. E-mail: dac@et.aau.dk

# Catalytic upgrading of hydrothermal liquefaction biocrudes: Different challenges for different feedstocks

Daniele Castello*, Muhammad Salman Haider, Lasse Aistrup Rosendahl

Department of Energy Technology, Aalborg University, Pontoppidanstræde 111, 9220 Aalborg Øst,

Denmark

*Corresponding author. E-mail: dac@et.aau.dk

**Abstract**

Hydrothermal liquefaction (HTL) followed by catalytic hydrotreating of the produced biocrude is increasingly gaining ground as an effective technology for the conversion of biomass into liquid biofuels. A strong advantage of HTL resides in its great flexibility towards the feedstock, since it is able to treat a large number of different organic substrates, ranging from dry to wet residual biomass. Nevertheless, the characteristics of biocrudes from different typologies of organic materials result in different challenges to be met during the hydrotreating step, leading to differences in heteroatoms removal and in the typology and composition of the targeted products. In this work, biocrudes were catalytically hydrotreated with a commercial NiMo/Al$_2$O$_3$ catalyst at different temperatures and pressures. Sewage sludge biocrude was found to be very promising for the production of straight-chain hydrocarbons in the diesel range, with considerable heteroatoms removal even at mild hydrotreating conditions. Similar results were shown by algal biocrude, although complete denitrogenation is challenging. Upgraded biocrudes from lignocellulosic feedstock (miscanthus) showed high yields in the gasoline range, with a remarkable content of aromatics. Operating at a higher H$_2$ pressure was found to be crucial to prevent coking and decarboxylation reactions.

**Keywords**

# 1. Introduction

Producing sustainable liquid fuels from renewable sources is one of the most exciting challenges of engineering nowadays. Liquid fuels are indeed utilized in very large amounts and, since they are mostly produced from fossil oil, they represent a remarkable contribution to global warming [1]. Although many steps forward have been made for the transition to more sustainable solutions for transportations, for example by means of a more extensive usage of electricity, liquid fuels still represent an important share of the world's energy demand. Indeed, applications such as long distance road transportations, aviation and marine traffic will rely on liquid fuels into the foreseeable, even long-term future. Such liquid fuels must be sustainable in all senses of the term, and biomass or organic waste streams are probably the most promising sources for the production of these [2]. Through the last decade, hydrothermal liquefaction (HTL) has been gaining a prominent position in the production of sustainable liquid fuels [3]. HTL consists in reacting biomass in hot pressurized water, at pressures high enough to keep water in its liquid or even supercritical state. The valuable product from this process is a dark viscous liquid, commonly referred to as biocrude, which is immiscible with water and which has a significantly higher heating value and lower oxygen content than bio-oils from other liquefaction technologies (e.g. pyrolysis). However, despite its remarkable characteristics, biocrude does not fulfill the requirements for drop-in transportation fuels due to the still high heteroatoms content, low hydrogen-to-carbon ratio and relatively high share of high-boiling fractions. Hence, upgrading processes are required, among which the most widely adopted one is hydrotreating, i.e. the reaction of biocrude with $H_2$ at high temperature and pressure in the presence of a catalyst [4,5].

In the literature, the number of studies concerning biocrude hydrotreating is quite limited. The large majority has been carried out on the upgrading of biocrudes from algal feedstock [6,7], for which the challenge is represented by the removal of nitrogen. The Pacific Northwest National Laboratory (USA) research group hydrotreated different algal biocrudes in a continuous bench-scale reactor [8,9], obtaining an almost complete removal of nitrogen, with a residual 1% of oxygen. Other studies in the literature were conducted with different catalysts and operating conditions, such as: ZSM-5 zeolites [10], sulfided NiMo and CoMo on $Al_2O_3$ support [11], $Pt/Al_2O_3$ and HZSM-5 [12] and several other supported metallic catalysts [13]. Two-stage upgrading has been investigated as well [14,15]. A few hydrotreating studies are also available for wood biocrude. Early work addressed the upgrading of wood biocrudes from the pioneering HTL experiments at the Albany PDU [5,16,17]. More recently, a parametric study on wood biocrude upgrading was carried out by Jensen et al. [18], showing that temperature is the most significant parameter affecting the extent of deoxygenation. The same author also reported the results of several hydrotreating campaigns performed by Steeper Energy [19] with different catalysts (NiMo/Al$_2$O$_3$, NiW/SiO$_2$/Al$_2$O$_3$, Pd/Al$_2$O$_3$). As far as waste biomass is concerned, a paper by Marrone et al. [20] investigated the hydrotreatment of biocrudes derived from primary sewage sludge and digested solids. Their results showed an almost complete denitrogenation but a residual presence of oxygen of around 1%.

The only work in the literature directly comparing upgraded biocrudes from different sources has been recently published by Jarvis et al. [21]. Here, the authors processed three upgraded biocrudes from different feedstocks (pine wood, microalga *Chlorella* and sewage sludge), by using a CoMo/Al$_2$O$_3$ catalyst at 400 °C and ca. 10.5 MPa. The differences among the upgraded oils were highlighted, finding that those derived from algae and sewage sludge had similar compositions but differed from upgraded pine biocrude, probably due to the higher content in lipids, proteins and cellulose of the former substrates. More in general, this work suggests that each biocrude has different key compounds and therefore it follows a different upgrading route. However, results were based on a single set of reaction conditions (400 °C and ~100 bar). It is reasonable to think that the choice of process parameters affects each biocrude in a different way, highlighting different reaction behaviors. These aspects still need to be investigated.

The present study aims at filling this gap, by testing different hydrotreating conditions on three biocrudes produced from as many feedstocks: miscanthus, microalga *Spirulina* and primary sewage sludge. These organic substrates represent biomasses of potential interest for HTL development, each of them showing particular characteristics. As a lignocellulosic biomass, miscanthus biocrude involves aromatic compounds deriving from lignin. Algal biomass, such as *Spirulina*, presents a high content of nitrogen-containing compounds. To a minor extent, this is also the case of primary sewage sludge, whose composition is however different due to a remarkable content of lipids. HTL biocrudes from these three feedstocks were processed at identical process conditions, testing different temperatures and H$_2$ pressures. The upgraded oils were fully characterized, highlighting the influence of process conditions on deoxygenation and denitrogenation efficiency and on the production of different fractional cuts. Results show that biocrudes from different feedstocks present different upgrading challenges and each of them is more indicated for the production of a different typology of potential drop-in fuel.

**2. Materials and methods**

*2.1. Materials*

Three different biocrudes were utilized in this study and were produced by Aarhus University, Denmark. The three biocrudes derive from the HTL of three different types of biomass: miscanthus, microalga *Spirulina* and primary sewage sludge from a municipal wastewater treatment plant. The choice of operating with these feedstocks was made in order to represent different typologies of biomass and hence to prove the high flexibility of this technology. HTL experiments were carried out in a continuous pilot plant operating at subcritical conditions (350 °C, 22 MPa). Details on the production of these biocrudes can be found in [22]. The received biocrudes were subjected to a filtration treatment, in order to reduce their ash content and hence allowing an easier way of processing. Filtration was carried out by dissolving each biocrude sample in acetone, then filtrating the resulting product and evaporating the solvent with a rotary evaporator. After this procedure, the resulting biocrudes were characterized by means of elemental analysis (see Section 2.3 for details).

Pre-sulfided Criterion Centera™ DN-3630 catalyst (NiMo/$Al_2O_3$) in the form of extrudates was utilized in the experimental runs.

*2.2. Experimental device and procedure*

Reactions were carried out in 25 mL autoclaves, constituted by a stainless steel tube sealed on the bottom. The top part of the reactors was connected to a capillary, closed by a ball valve, allowing the injection and sampling of gases. A picture of the reactors can be found in [18]. Before each experiment, the reactor was loaded with 4 g of biocrude and 2 g of NiMo/$Al_2O_3$ catalyst, hence obtaining a catalyst-to-oil ratio of 0.5. Three stainless steel spheres were also loaded in each reactor, in order to enhance mixing during the operations. The reactors were then sealed, purged and pressurized with $H_2$ to the desired initial pressure ($P_0$) according to the experiment. The bottom part of the reactors was then inserted in a fluidized sand bath SBL-2D (Techne, UK) heated at the desired reaction temperature, while the top part was connected to a mechanical agitator, providing the necessary mixing throughout the desired reaction time. Reaction pressure was continuously monitored and logged by means of a pressure transducer Wika A-10 interfaced with a LabView™ software.

All reactions were run for 4 h. The choice of the reaction time and the amounts of biocrude and catalyst was operated in order to mimic a continuous hydrotreater operated at a weight hourly space velocity (WHSV) of 0.5 $h^{-1}$, which is in the typical range of conditions for these type of process [9,23]. This parameter implies that, in a continuous reactor, 1 g of catalyst contacts 0.5 g of feed every hour. The same ratio was achieved in the present study.

After reaction, the reactors were quenched in water and gases were collected in a sealed tube for further analyses. Then, the reactors were opened and the liquid products were passed through a metallic mesh in order to separate the catalyst, while the liquid products (i.e. oil and water phases) were collected and weighed. The vial containing the liquid product was then centrifuged at 4,000 rpm for 5 minutes, in order to separate the oil phase and the water phase. The interior of the reactor and the catalyst itself were accurately washed with around 100 mL of acetone. Solids (catalyst and coke) were separated over a paper filter and then dried at 120 °C overnight, in order to determine the amount of coke by weight difference. Acetone was removed from the liquid filtrate in a rotary evaporator at 550 mbar and 60 °C. Oil yield was thus given by the sum of the oil recovered from direct collection and the one recovered after acetone washing. However, the oil sample that underwent analysis was only the one recovered in the first step, i.e. not subjected to solvent extraction, in order to avoid any possible alteration of its composition.

For each biocrude, three experiments were performed at different temperatures (350±5 and 400±5 °C) and initial pressures $H_2$ (4.0±0.2 and 8.0±0.4 MPa). A complete list of the test runs is reported in Table 1, along with the average pressure measured at reaction conditions ($P_r$). In order to ascertain reproducibility, each test run was performed in duplicates by using two identical reactors. Experimental results are reported as the mean value of the repeated experiments, with an error represented by their standard deviation. The experimental error for derived measures was calculated by means of error propagation.

*Table 1* – *Summary of the performed hydrotreating experiments.*

| Experiment | Biocrude | Temperature (°C) | Res. Time (h) | $P_0$ (MPa) | $P_r$ (MPa) |
|---|---|---|---|---|---|
| AL350-4 | *Spirulina* | 350 | 4 | 4.0 | 9.9 |
| AL350-8 | *Spirulina* | 350 | 4 | 8.0 | 12.3 |
| AL400-8 | *Spirulina* | 400 | 4 | 8.0 | 15.6 |
| SS350-4 | Sewage sludge | 350 | 4 | 4.0 | 9.2 |
| SS350-8 | Sewage sludge | 350 | 4 | 8.0 | 14.0 |
| SS400-8 | Sewage sludge | 400 | 4 | 8.0 | 15.7 |
| MS350-4 | Miscanthus | 350 | 4 | 4.0 | 10.8 |
| MS350-8 | Miscanthus | 350 | 4 | 8.0 | 15.6 |
| MS400-8 | Miscanthus | 400 | 4 | 8.0 | 16.9 |

*2.3. Analytic techniques*

Reaction products, as well as the initial feeds, were characterized by means of several analytic techniques. The gaseous products were analyzed by means of a GC-2010 gas chromatograph with a barrier ionization discharge detector (GC-BID), manufactured by Shimadzu Inc. and equipped with a Supelco 1006 PLOT column. The measured $H_2$ concentration in the gas was utilized to determine hydrogen consumption, known the initial and final gas pressures in the reactor. Ideal gas law was utilized to this purpose.

The elemental composition of the raw and upgraded biocrudes in terms of C, H and N was measured by a Perkin Elmer CHN-O analyzer model 2400 Series II, following the standard ASTM D5291 [24]. The concentration of O was calculated by difference. The elemental composition of the oil served as a base for the calculation of the degree of deoxygenation (de-O) and denitrogenation (de-N), which were calculated as follows:

$$\text{de-O} = 1 - \frac{O_{product}}{O_{feed}} \tag{1}$$

$$\text{de-N} = 1 - \frac{N_{product}}{N_{feed}} \tag{2}$$

The elemental composition of both raw and upgraded biocrudes was also adopted for the calculation of the high heating value (HHV), which was estimated by using the Channiwala-Parikh correlation [25].

The distillation curves of both raw and upgraded biocrudes were determined by means of simulated distillation (SimDis). The equipment used to this purpose was a gaschromatograph GC-2010 (Shimadzu Inc., Japan), with a flame ionization detector (FID) and a Zebron ZB-1XT column (Phenomenex, Germany). SimDis analysis was carried out according to the standard ASTM D7169 [26]. However, in order to provide better solubility for the analyzed samples, dichloromethane (DCM) was utilized instead of $CS_2$, following the approach outlined by Vardon et al. [27].

The oil samples were also analyzed by gas chromatography with mass spectrometry (GC-MS) Trace 1300 ISQ QD – Single Quadrupole (Thermo Scientific), equipped with a HP-5MS column. Samples were diluted at 1% in diethyl-ether (DEE) and injected at 300 °C with a split ratio of 1:20. Helium was used as carrier gas, with a flow rate of 1 mL/min. Measuring program involved heating at 10 °C/min from 35 °C to 120 °C, holding the temperature for 5 min and then 5 °C/min up to 250 °C. Spectra were interpreted by means of NIST libraries. Relative peak areas were calculated by considering the 100 peaks with the largest area, excluding the solvent one.

## 3. Results and discussion

In this section, the results of the different tests are presented. First, the removal of oxygen and nitrogen heteroatoms is discussed (Section 3.1). The different oils are then compared in terms of boiling point distribution, in order to assess the potential production of fuel cuts (Section 3.2). Finally, the chemical composition of raw and upgraded biocrudes is discussed by means of GC-MS measurements (Section 3.3).

*3.1 Heteroatoms removal, carbon loss and $H_2$ consumption*

The removal of heteroatoms represents the most demanding aspect on biocrude upgrading. In Table 2 it can be clearly observed how the upgraded oils present a reduced content of both oxygen and nitrogen compared to the raw ones. The removal of heteroatoms also results in an increased HHV of the upgraded oils, especially as a consequence of their lower oxygen content. None of the samples showed a complete removal of both oxygen and nitrogen, although for some of them complete deoxygenation was achieved, at least within the error margin of CHN measurements.

*Table 2* – *Yields, elemental composition, H/C ratio and HHV of the analyzed oil samples. Oxygen was calculated by difference.*

| Experiment  | Yield (wt. %) | Elemental composition (wt. %) | | | | H/C (-) | HHV (MJ/kg) |
|---|---|---|---|---|---|---|---|
|  |  | C | H | N | O |  |  |
| AL biocrude | - | 75.0 ± 0.3 | 10.4 ± 0.1 | 7.7 ± 0.1 | 6.9 ± 0.3 | 1.66 ± 0.02 | 37.7 ± 0.2 |
| AL350-4 | 75 ± 3 | 82.2 ± 0.2 | 11.1 ± 0.1 | 5.4 ± 0.0 | 1.3 ± 0.2 | 1.62 ± 0.02 | 41.6 ± 0.1 |
| AL350-8 | 73 ± 2 | 84.0 ± 0.4 | 12.1 ± 0.1 | 4.0 ± 0.2 | 0.0 ± 0.5 | 1.72 ± 0.02 | 43.5 ± 0.2 |
| AL400-8 | 63 ± 4 | 83.7 ± 0.4 | 12.3 ± 0.2 | 4.1 ± 0.2 | 0.0 ± 0.5 | 1.76 ± 0.03 | 43.7 ± 0.3 |
|  |  |  |  |  |  |  |  |
| SS biocrude | - | 74.5 ± 0.1 | 10.6 ± 0.0 | 3.9 ± 0.0 | 11.0 ± 0.1 | 1.71 ± 0.00 | 37.4 ± 0.0 |
| SS350-4 | 74 ± 3 | 83.1 ± 0.7 | 12.1 ± 0.1 | 3.6 ± 0.0 | 1.2 ± 0.7 | 1.75 ± 0.02 | 43.1 ± 0.3 |
| SS350-8 | 77 ± 2 | 84.1 ± 0.4 | 13.4 ± 0.0 | 2.5 ± 0.0 | 0.0 ± 0.4 | 1.91 ± 0.01 | 45.1 ± 0.1 |
| SS400-8 | 72 ± 3 | 85.3 ± 0.2 | 13.8 ± 0.0 | 0.9 ± 0.0 | 0.0 ± 0.2 | 1.95 ± 0.00 | 46.1 ± 0.1 |
|  |  |  |  |  |  |  |  |
| MS biocrude | - | 70.5 ± 0.6 | 8.2 ± 0.0 | 1.7 ± 0.1 | 19.6 ± 0.6 | 1.40 ± 0.01 | 32.2 ± 0.2 |
| MS350-4 | 60 ± 4 | 81.1 ± 0.3 | 8.8 ± 0.0 | 1.1 ± 0.1 | 9.0 ± 0.3 | 1.30 ± 0.00 | 37.8 ± 0.1 |
| MS350-8 | 65 ± 1 | 84.7 ± 0.2 | 10.2 ± 0.1 | 0.8 ± 0.0 | 4.3 ± 0.2 | 1.45 ± 0.01 | 41.1 ± 0.1 |
| MS400-8 | 61 ± 3 | 87.4 ± 0.0 | 10.3 ± 0.3 | 1.5 ± 0.1 | 0.8 ± 0.3 | 1.37 ± 0.04 | 42.2 ± 0.4 |

Heteroatoms removal causes a reduction in the yields of upgraded oil, as O and N are converted into gaseous and other liquid products. However, the yields of biocrude do not directly follow the extent of heteroatoms removal. Indeed, it can be appreciated that the oil yields of experiments at 350 °C - 8 MPa are comparable or even higher to those at 350 °C - 4 MPa, although the former experiments remove more heteroatoms. This is explained with the lower availability of hydrogen when operating at low pressures, resulting in coking (especially for unstable feedstocks) and more extensive gas formation. Operating at higher temperature results also in reduced yields, due to increased heteroatoms removal but, also, to a higher extent of cracking and coking reactions. These effects can be observed in Table 3, where the carbon balances are reported. The way reaction conditions affect carbon loss for the different biomasses considerably differs. For *Spirulina*, the most unfavorable conditions are those at high temperature, at which, probably, nitrogen-containing compounds are unstable, despite the high pressure. On the contrary, miscanthus biocrude gave the highest amounts of coke at low pressure, therefore when the available hydrogen was not sufficient to stabilize the oil. In all cases, only minor carbon loss was associated with decarboxylation and decarbonylation. On the other hand, it should be pointed out that CO and $CO_2$ can in turn be converted into hydrocarbons and water through other gas-phase reactions (e.g. water-gas shift and CO methanation).

*Table 3* – *Carbon balance of the performed experiments. Carbon yields are expressed in wt. % with respect to carbon in the feed. Coke was assumed to be composed of pure C. C in the aqueous phase was not quantified.*

| Experiment | Oil (wt. C %) | Gas (wt. C %) | | Coke (wt. C %) | Total (wt. C %) |
|---|---|---|---|---|---|
| | | $CO-CO_2$ | Hydrocarbons | | |
| AL350-4 | 82.2 | 0.4 | 2.0 | 2.1 | 86.8 |
| AL350-8 | 81.8 | 0.0 | 0.7 | 2.5 | 85.0 |
| AL400-8 | 70.4 | 0.0 | 1.4 | 9.3 | 81.2 |
| SS350-4 | 83.1 | 1.0 | 2.3 | 4.2 | 90.0 |
| SS350-8 | 86.9 | 0.7 | 2.7 | 2.0 | 92.3 |
| SS400-8 | 82.4 | 0.3 | 1.7 | 3.1 | 87.4 |
| MS350-4 | 69.0 | 0.4 | 4.0 | 14.1 | 87.5 |
| MS350-8 | 78.1 | 0.1 | 1.2 | 5.8 | 85.2 |
| MS400-8 | 75.6 | 0.4 | 3.6 | 5.4 | 85.0 |

An interesting point is represented by the analysis of heteroatoms removal and the necessary uptake of hydrogen, which is shown in Figure 1. It can be visually appreciated how deoxygenation is relatively easier to achieve than denitrogenation. For sewage sludge and *Spirulina*, even mild conditions at 350 °C and 4 MPa are able to give a deoxygenation efficiency higher than 80%. Reactions conducted at $P_0 = 8$ MPa are then able to obtain complete deoxygenation. The extent of denitrogenation is definitely lower. For algal biocrude, denitrogenation goes from around 30% at 350 °C and 4 MPa to no more than 48% when operating at 8 MPa, regardless from temperature. It should be remarked that high nitrogen removal is not achieved in the literature, with the only significant exception of results from PNNL, which were obtained in a continuous device [8,9]. The occurrence of a better mass transfer is thought to be the cause of the better performance of that study [11]. The trend of heteroatoms removal, in particular denitrogenation, is reflected by that of $H_2$ consumption. Indeed, $H_2$ consumption is relatively low (6 mol/kg$_{biocrude}$) at mild conditions (low de-N and de-O) and it increases to around 14 mol/kg in the experiment at 8 MPa, where a higher denitrogenation is achieved. $H_2$ consumption remains approximately constant at 400 °C, where the same de-O and de-N were obtained.

*Figure 1* – *Degree of de-oxygenation and de-nitrogenation (left axis) and $H_2$ consumption (right axis) observed in the hydrotreating experiments.*

For sewage sludge biocrude, the removal of oxygen revealed to be very high even at mild conditions, while denitrogenation increases with reaction severity. As with algal biocrude, the consumption of $H_2$ generally follows the trend of denitrogenation but with a significant exception. Indeed, when processing at 400 °C, similar $H_2$ consumption as at 350 °C is obtained, but the extent of heteroatoms removal is significantly higher. A possible explanation could be the occurrence of high temperature reactions, able to remove oxygen or nitrogen without the utilization of $H_2$. However, another explanation could be found in the gas-phase reactions involving CO, $CO_2$ and $H_2$. During hydrotreating, CO and $CO_2$ are produced from the decarbonylation and decarboxylation of the oxygenates. These gases can react with $H_2$ to eventually form $CH_4$ via the reverse water-gas shift and CO methanation reactions, generating an extra $H_2$ consumption [28,29]. Hence, minimizing decarboxylation and decarbonylation may result in a higher availability of $H_2$. In Section 3.3 evidence of less decarboxylation occurring at more severe conditions is given. Further research is however needed to understand this phenomenon.

Miscanthus showed also an increase in heteroatoms removal from 4 MPa to 8 MPa. However, when temperature is raised to 400 °C, deoxygenation improves but, surprisingly, a lower degree of denitrogenation is achieved, although the difference between N concentration is not very large in absolute terms (0.8 % vs. 1.5%). This can be caused by the onset of competition between deoxygenation and denitrogenation for the utilization of $H_2$. Reaction rates for deoxygenation appear much more accelerated than those for denitrogenation, which makes them able to utilize $H_2$. Such phenomenon can be more evident for miscanthus due to its much higher initial oxygen content.

Apart from heteroatoms removal, hydrogen is also employed for the hydrogenation of the oils, i.e. the saturation of C-H bonds, leading to a more aliphatic nature of the fuel. Upgraded oils generally show an increase in the H/C ratio. This is true for algal and sewage sludge oils, as it can be observed in Table 2. For sewage sludge oils in particular, the hydrotreatment at 400 °C leads to a H/C ratio of 1.95, which reveals its highly paraffinic nature. Oppositely, miscanthus oils show negligible or even negative variations of the H/C, especially at higher temperature. Low values of H/C are indicative of the highly aromatic nature of these oils, in which the phenolic compounds deriving from lignin are playing an important role (Section 3.3). Moreover, miscanthus biocrudes show the highest initial heteroatoms content. It can be deduced that, at the investigated reaction conditions, $H_2$ is primarily used for heteroatoms removal rather than for hydrogenation.

*3.2 Boiling point distribution and fractional cuts*

Boiling point distribution is an important aspect to assess the drop-in potential of a fuel. Additionally, it allows observing the extent of cracking or polymerization reaction during the upgrading process. In Figure 2, the boiling point distributions obtained by simulated distillation are reported.

***Figure 2*** – *Boiling point distribution determined by SimDis (ASTM D7169) for (a) Spirulina, (b) sewage sludge and (c) miscanthus raw and hydrotreated biocrudes.*

A first aspect that can be deduced from the curves is that the three investigated biocrudes are very different one from each other. In particular, miscanthus biocrude appears much heavier than the other two. This is evident from the lower value of recovery shown by this oil (65%), compared to *Spirulina* (86%) and sewage sludge biocrude (93%). This aspect reveals the presence in this oil of compounds with a much higher molecular weight than the other two biocrudes, most likely represented by lignin-derived polymers.

After hydrotreating, SimDis curves usually show a shift towards lower boiling points. The entity of this shift follows the severity of the adopted hydrotreating conditions. Conducting reactions at high temperature and pressure enhances cracking reactions, which form lower boiling point compounds. However, it must be also pointed out that the Sim-Dis technique is calibrated with a mixture of paraffins, hence the results for oxygenated compounds are not completely accurate, due to the increase in the boiling point of the oxygenates with respect to the corresponding hydrocarbons [27].

Tests performed at a pressure of 4 MPa do not show significant changes compared to raw biocrude. For miscanthus, only moderate cracking can be observed between 200-300 °C, while for the other two biocrudes, the hydrotreated liquid presents characteristics even worse than the starting biocrude. This effect is particularly evident for sewage sludge oil (SS350-4), showing a lower value of recovery compared to its raw biocrude. A closer observation of the SimDis curve (Figure 2b) reveals an evident decrease in the recovery between 300-400 °C, corresponding to molecules in the range C20-30. To a lower extent, this is also the case of the *Spirulina* oil at the same conditions (AL350-4). Operating at low pressure conditions, hence with less available $H_2$, exposes the oils to possible polymerization and coking reactions. This circumstance appears as more likely as higher is the oxygen content and the presence of compounds prone to polymerization. Operating at higher $H_2$ pressures is therefore beneficial especially for highly unstable biocrudes.

When higher pressures are adopted, a noticeable increase in the distribution is observed, i.e. high recoveries at the same boiling point are found. The curves obtained at 350 °C and 8 MPa generally show similar recoveries than the raw biocrudes at low temperatures below 200 °C, while at higher temperatures they start to considerably diverge. The effect of cracking is therefore especially expressed in the boiling range of light diesel, as it can be observed in Table 4.

*Table 4* – *Fractional cuts obtained from the simulated distillation of the raw and upgraded biocrude samples. Values are reported in wt. % recovery.*

| Sample | Gasoline (< 193 °C) | Jet-fuel (193-271 °C) | Light diesel (272-321 °C) | Heavy diesel (321-425 °C) | VGO (425-564 °C) | Residue (> 564 °C) |
|---|---|---|---|---|---|---|
| AL biocrude | 16.2 | 11.7 | 8.5 | 25.2 | 15.9 | 22.5 |
| AL350-4 | 7.1 | 16.3 | 13.7 | 20.4 | 19.2 | 23.3 |
| AL350-8 | 17.8 | 23.0 | 20.9 | 16.2 | 10.4 | 11.7 |
| AL400-8 | 17.8 | 26.5 | 22.2 | 14.1 | 7.1 | 12.3 |
| SS biocrude | 5.5 | 12.8 | 14.1 | 37.6 | 14.5 | 15.5 |
| SS350-4 | 6.9 | 11.9 | 20.1 | 17.2 | 16.8 | 27.1 |
| SS350-8 | 11.3 | 17.1 | 31.2 | 15.9 | 12.3 | 12.2 |
| SS400-8 | 14.1 | 20.6 | 31.4 | 14.2 | 8.3 | 11.4 |
| MS biocrude | 11.3 | 8.1 | 14.5 | 13.1 | 11.7 | 41.3 |
| MS350-4 | 13.1 | 13.7 | 8.3 | 12.8 | 12.8 | 39.3 |
| MS350-8 | 13.4 | 16.1 | 11.9 | 17.2 | 15.0 | 26.4 |
| MS400-8 | 26.9 | 20.7 | 11.8 | 14.0 | 9.3 | 17.3 |

Increasing the hydrotreating temperature to 400 °C induces further cracking in the upgraded oils, although this effect seems to be advantageous especially for the low boiling point fractions. Temperature increase results in higher jet-fuel and gasoline production, while diesel yields remain approximately unchanged. Operating at higher temperatures thus helps in the conversion of the heavy ends into useful drop-in products, with enhanced heteroatoms removal. On the other hand, as it was discussed in Section 3.1, operating at higher temperatures generally results into lower yields. A correct balance between these contrasting tendencies should be found in actual applications.

The amounts of different fractional cuts that can be obtained vary with the feedstock. Among the three biocrudes, upgraded miscanthus oil (MS400-8) showed the highest relative concentration of gasoline (26.9%), witnessing the presence of relatively light compounds, up to $C_{12}$. The production of gasoline is strongly influenced by the temperature and it is more than doubled when going from 350 to 400 °C. The highest relative production of light diesel is obtained by sewage sludge oil (SS400-8), with 31.4%. *Spirulina* biocrude (AL400-8) showed, instead, the highest fraction of jet-fuel (26.5%), with results in line with the findings of Biller et al. [11]. However, it should be pointed out that the set of conditions at which HTL is conducted is likely to play an important role, as it can determine the amount of heteroatoms and the molecular weight distribution of the produced biocrude. In this work, the three biocrudes were produced at the same HTL operating conditions [22], but it is reasonable to expect that an optimized set of HTL conditions, specific for each biomass, would have a strong impact on the potential production of specific drop-in fuel cuts [3]. Moreover, optimizing the hydrotreating parameters can lead to significantly different results. For example, significant production of middle distillates from lignocellulosic biocrudes was reported by Jensen et al. [19] after a two-stage upgrading process.

*3.3 Composition of the volatile fraction of the oils*

A closer look can be given to the volatile fraction of the raw and upgraded biocrudes by means of GC-MS analysis. Results from GC-MS are only limited to the volatile fraction of the sample, i.e. up to a boiling point of ca. 350 °C. Nevertheless, this fraction is the one more directly involved in the production of potential drop-in fuels, therefore it is possible to evaluate the nature of the compounds produced from the process. Additionally, hypotheses on reaction pathways can be formulated, understanding the mechanisms involved during hydrotreatment.

In Figure 3, results of GC-MS analysis are reported in terms of families of compounds, by analyzing the 100 peaks with the largest area in each chromatogram. In particular, the different classes of hydrocarbons were discriminated, along with oxygen- and nitrogen-containing compounds.

**Figure 3** – *Composition of the analyzed oils by means of GC-MS in terms of families of compounds.*

A noticeable aspect is the dramatic change in chemical composition following hydrotreating. The composition of untreated biocrudes is indeed dominated by oxygenated and nitrogenated species, which sum up to 83.7% for sewage sludge, 68.7% for algae and 82.0% for miscanthus biocrudes. Hydrocarbons are almost absent from raw biocrudes, with the exception of *Spirulina* biocrude, where 11.6% of the detected compounds can be classified as *n*-paraffins. After hydrotreating, the composition of the biocrudes visibly changes. The amount of oxygen- and nitrogen-containing compounds decreases as reaction conditions become more severe, while hydrocarbons are formed. The relative amount of hydrocarbons in the upgraded oils ranges from 40% obtained at mild conditions to more than 90%. However, strong differences are shown by the different biocrudes both for the extent of upgrading and for the types of compounds found in the crude.

*3.3.1 Key-compounds and reactions routes*

The composition of upgraded biocrudes from *Spirulina* consists of around 50% of *n*-paraffinic hydrocarbons. This amount remains approximately constant with the increase of reaction severity. At higher $H_2$ pressures, more hydrocarbons are formed, but especially in the form of iso-paraffins, naphthenes and aromatics. The content of oxygen- or nitrogen-containing compounds is reduced as far as reaction conditions become more severe. While in the raw biocrude the majority of the peak area (ca. 69%) is represented by heteroatom-containing compounds, this percentage is reduced to around 10% in the AL400-8 upgraded oil. As a consequence, the share of peak area represented by hydrocarbons goes from 15% in the raw biocrude to 78% in AL400-8 oil.

The observation of the GC-MS chromatograms of both raw and upgraded biocrudes (Figure 4) can help understanding the ongoing phenomena. Raw biocrude shows an elevated number of different compounds, belonging to different chemical families. An evident peak is represented by *n*-$C_{17}$, which is found in appreciable concentrations in the biocrude, derived from the decarboxylation of linolenic acid, which is an important constituent of this type of algae [30]. More to the right, a number of peaks represent compounds with long aliphatic chains and different types of substituting groups. For example, at elution times between 35 and 40 minutes, a group of $C_{12}$ amides is present. It can be observed that none of these peaks is then found in the hydrotreated oil, most likely being converted into paraffinic hydrocarbons (Figure 5). It can be postulated that the same mechanism takes place for longer aliphatic chains, which cannot be detected by the present system due to their high boiling point. Nevertheless, other studies in the literature confirm the presence of these larger amides, such as hexadecanamide and octadecanamide [11].

**Figure 4** – *Comparison of the GC-MS chromatograms for raw and upgraded biocrude from Spirulina.*

The left part of the chromatogram shows smaller molecules, many of them with an aromatic structure, such as p-cresol and other phenolic compounds. Additionally, heterocyclic nitrogen-containing compounds are detected. The most abundant species in this class of molecules are indole and pyrrole, with many different alkyl substituents. Indole and pyrrole, as well as other heterocyclic nitrogenated compounds (e.g. pyridine and quinoline), are originated from the Maillard reaction, involving carbohydrates and nitrogen-containing compounds [31]. Compared to amides, amines or nitriles, heterocyclic nitrogen is more difficult to remove, requiring more $H_2$ and more severe reaction conditions [32]. Indole conversion was also observed in the reported set of experiments. Based also on literature [15], possible reaction products can be represented by ethylbenzene and ethylcyclohexane (Figure 5), which were both detected among the reaction products.

*Figure 5 – Hydrodenitrogenation routes for some representative nitrogen-containing compounds.*

Compared to *Spirulina*, sewage sludge biocrude has a much lower nitrogen content. The most relevant constituents of this biocrude are fatty acids with an even number of carbon atoms, such as hexadecanoic (palmitic) acid, tetradecanoic (myristic) acid and dodecanoic (lauric) acid. This composition is confirmed by other studies [21,22], where different settings of the instruments were able to detect also other fatty acids up to $C_{20}$.

This type of compounds is relatively easy to upgrade into the corresponding straight-chain paraffins, even at mild conditions (Figure 3) [33]. This fact can be evidently observed in the chromatogram of the upgraded oil, which looks much less complicated and with clear peaks at almost regular elution time intervals, represented by normal paraffins. The most represented peaks are $C_{15}$-$C_{18}$, corresponding to the field of light diesel. Unlike *Spirulina*, the portion of the spectrum at the lowest elution times (i.e. in the gasoline range) is less populated and nitrogen-containing compounds can be hardly found, although they were detected by CHN analysis (Table 2). It can be hypothesized that such compounds are mostly present in higher boiling point fractions, thus not being detectable via GC-MS. This assumption has been recently proved by our group for algal biocrudes [34], but its validity to sewage sludge still needs to be confirmed.

*Figure 6 – Comparison of the GC-MS chromatograms for raw and upgraded biocrude from sewage sludge.*

The severity of the hydrotreating conditions has a strong impact in selecting the upgrading pathways. This aspect can be appreciated by observing the distribution of the carbon numbers in the upgraded oil. In Figure 7, the relative peak areas of the most represented hydrocarbons (*n*-$C_{15}$ – *n*-$C_{18}$) are shown.

*Figure 7* – *Relative peak areas of the most representative alkanes in the upgrading experiments with sewage sludge biocrude.*

When operating at mild conditions, hydrocarbons with an odd number of carbon atoms are found in higher amounts than even-numbered ones. As far as conditions become more severe, even-numbered hydrocarbons become predominant. This aspect reveals the occurrence of two reaction pathways for the deoxygenation of fatty acids, which are the most abundant class of oxygenates in primary sewage sludge biocrude. As it has been shown, sewage sludge biocrudes show different fatty acids with an even number of carbon atoms, e.g. tetradecanoic acid or hexadecanoic acid. Deoxygenation can proceed through hydrodeoxygenation, i.e. with the removal of oxygen in the form of $H_2O$. In this case, the total number of carbon atoms is preserved, leading to the formation of an even-numbered hydrocarbon. However, when $H_2$ is not available in large amounts, decarboxylation takes place. In this case, carboxylic group is converted into $CO_2$, thus removing one C atom and leading to an odd-numbered hydrocarbon (Figure 8) [35]. The occurrence of decarboxylation at less severe reaction conditions can be also appreciated in Table 3 and finds its conformation in several literature studies [36,37]. Although decarboxylation is effective as a deoxygenation strategy, it also causes carbon losses and thus it should be avoided, as it negatively affects oil yields.

*Figure 8* – *Reaction pathways for the deoxygenation of hexadecanoic acid into n-hexadecane (hydrodeoxygenation) and n-pentadecane (decarboxylation).*

*Spirulina* and sewage sludge biocrudes include a large portion of long-chain aliphatic components, due to the presence of fatty acids and amides. As a lignocellulosic biomass, miscanthus biocrude composition is strongly different. Moreover, compared to the other two biocrudes, it presents a higher oxygen concentration. From Figure 3, it can be appreciated how the adoption of more severe reaction conditions progressively leads to a lower extent of oxygenated compounds and an increasing share of hydrocarbons. However, compared to the other two oils, in which mostly paraffinic hydrocarbons were produced, in miscanthus naphthenic and aromatic hydrocarbons are produced in larger amounts, especially at high temperature conditions.

The observation of the chromatogram in Figure 9 makes this difference visible. Unlike the other oils, miscanthus biocrude presents many peaks in the range 5-15 min, due to small oxygenated compounds such as ketones and phenols. The former, mainly found in the form of alkylated cyclopentenones, are generated from the hydrothermal reactions of carbohydrates, while the latter are mainly derived from lignin, although some phenols are also generated from carbohydrates [38].

*Figure 9* – *Comparison of the GC-MS chromatograms for raw and upgraded biocrude from miscanthus.*

The uprading of miscanthus biocrude resulted in the production of a number of compounds derived from the hydrodeoxygenation and hydrogenation of the HTL products. In particular, two classes of hydrocarbons can be found: naphthenic (i.e. cycloalkanes) and aromatics. In the first group, the most found compounds are cyclopentanes and cyclohexanes, with different alkyl substituents. In the field of aromatics, a number of species involving a benzene ring are formed, as well as indene and naphthalene. These aromatic compounds are likely to derive from the corresponding phenols through hydrodeoxygenation. Through further hydrogenation, cyclohexanes can be produced. Cyclopentanes can be instead considered the upgrading products from cyclopentenones. Although most of the chromatogram peaks are found in the low boiling point region, $n$-$C_{15}$ - $n$-$C_{18}$ paraffins can be clearly noticed. This testifies the ongoing of upgrading reactions involving compounds in the heavy ends of the biocrude. As it was highlighted by Pedersen et al. [39], molecules with such a number of carbon atoms can be found in the distillation residue of a ligno-cellulosic biocrude. Among these compounds, straight-chain $C_{16}$-$C_{20}$ fatty acids can be found, although most species feature one or more aromatic rings with different alkyl chains. Enhancing the production of middle distillates would require effective deoxygenation and hydrogenation of these larger compounds.

## 4. Conclusions

An outstanding advantage of HTL resides in its flexibility towards the feedstock. However, it should be pointed out that biocrudes from different sources can require quite different upgrading strategies and are suitable for different types of final products. The issue of heteroatoms removal was successfully addressed with the use of a standard NiMo/$Al_2O_3$ sulfided catalyst, achieving complete deoxygenation and an appreciable degree of denitrogenation for biocrudes from *Spirulina* and sewage sludge. Sewage sludge revealed to be a highly promising feedstock for drop-in biofuels, with a remarkable production of straight-chain paraffinic hydrocarbons in the light diesel range. Significant production of hydrocarbons in both jet-fuel and diesel ranges was also obtained from microalga *Spirulina,* although with slightly higher aromaticity. Upgraded lignocellulosic biocrude from miscanthus appeared sensibly different from the other two, resulting in the production of compounds in the gasoline range with a remarkable content of aromatics. Operating with considerable amounts of $H_2$ at high pressures is crucial in order to avoid unwanted decarboxylation reactions and coking, both reducing the yields and potentially causing problems during operations in continuous plants. High temperatures are generally beneficial to the process, although the occurrence of overcracking must be carefully evaluated.

While complete oxygen removal can be readily achieved, the removal of nitrogen still represents a relevant challenge to be addressed. The effectiveness of nitrogen removal depends on the specific feedstock, being it a function of the specific nitrogen-containing compounds found in each biocrude. Strategies for nitrogen removal should be implemented, including the design of tailor-made catalysts and an effective selection of HTL conditions to minimize heterocyclic compounds.

**Acknowledgments**

This research has received funding from the European Union's Horizon 2020 Research and Innovation Program under grant agreement no. 764734 (HyFlexFuel). The funding source had no involvement in study design; in the collection, analysis and interpretation of data; in the writing of the report; and in the decision to submit the article for publication.

FIGURE 1

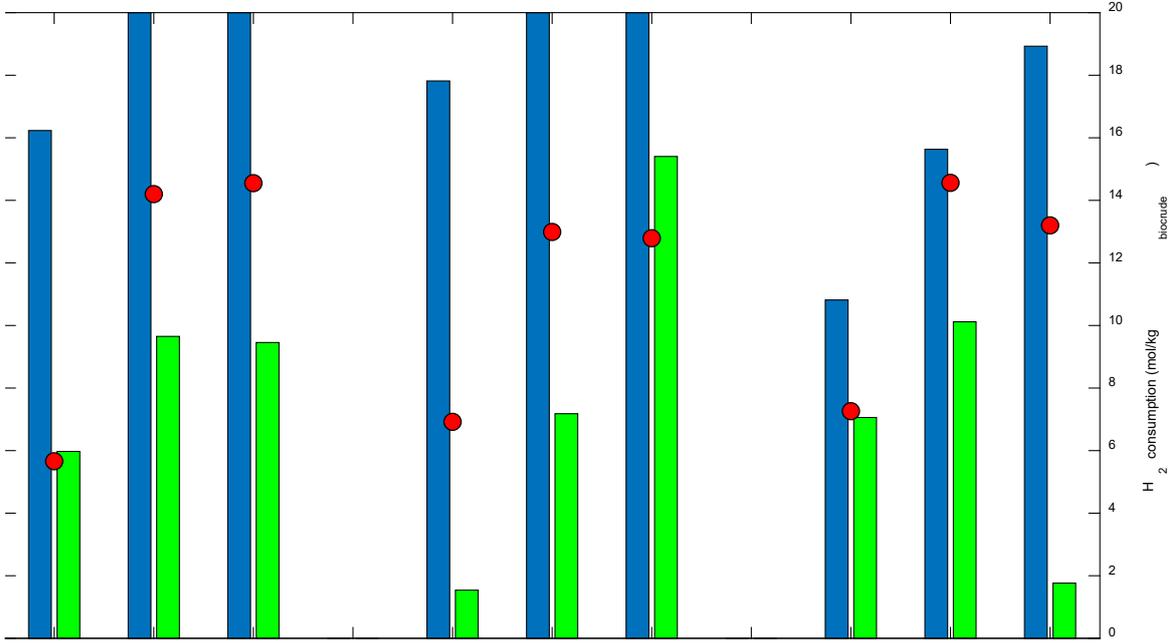

FIGURE 2

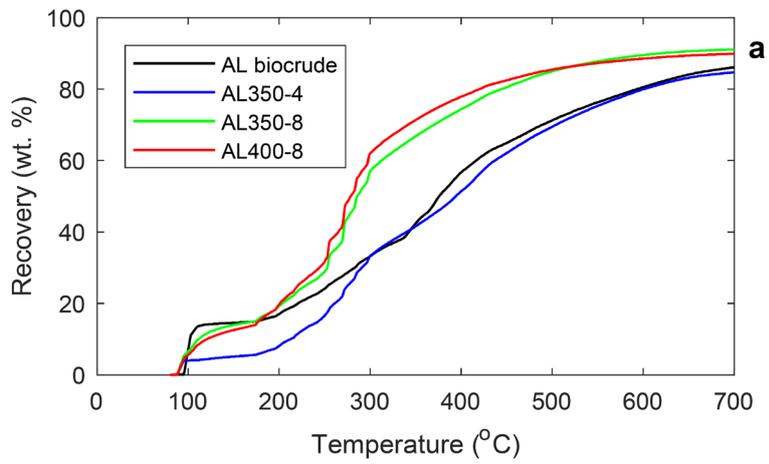

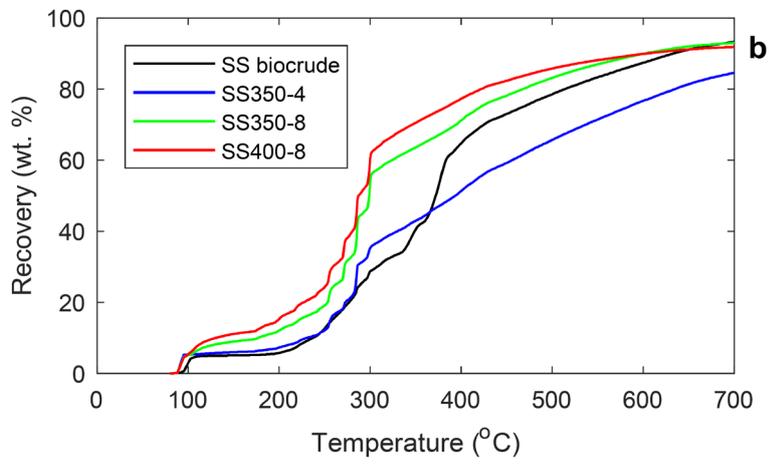

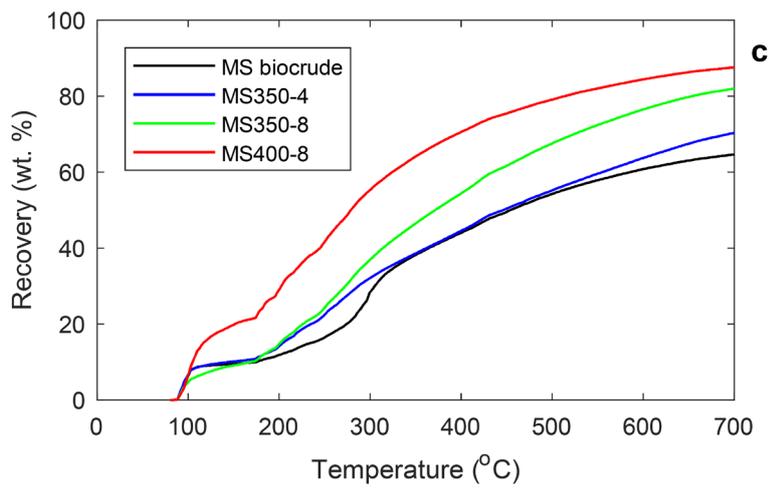

FIGURE 3

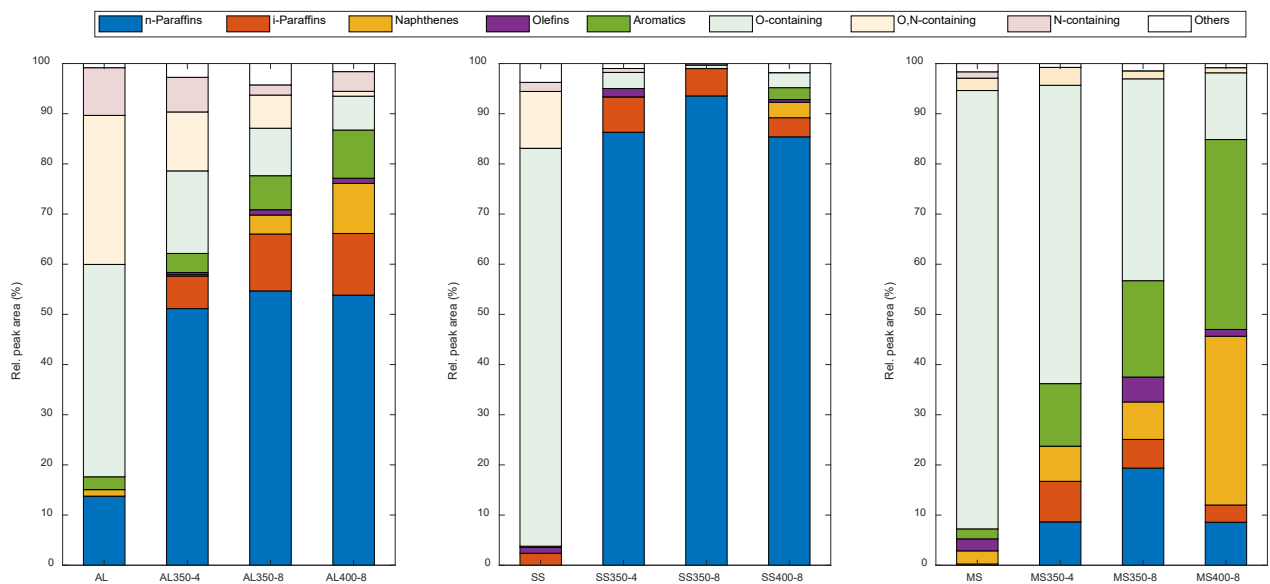

FIGURE 4

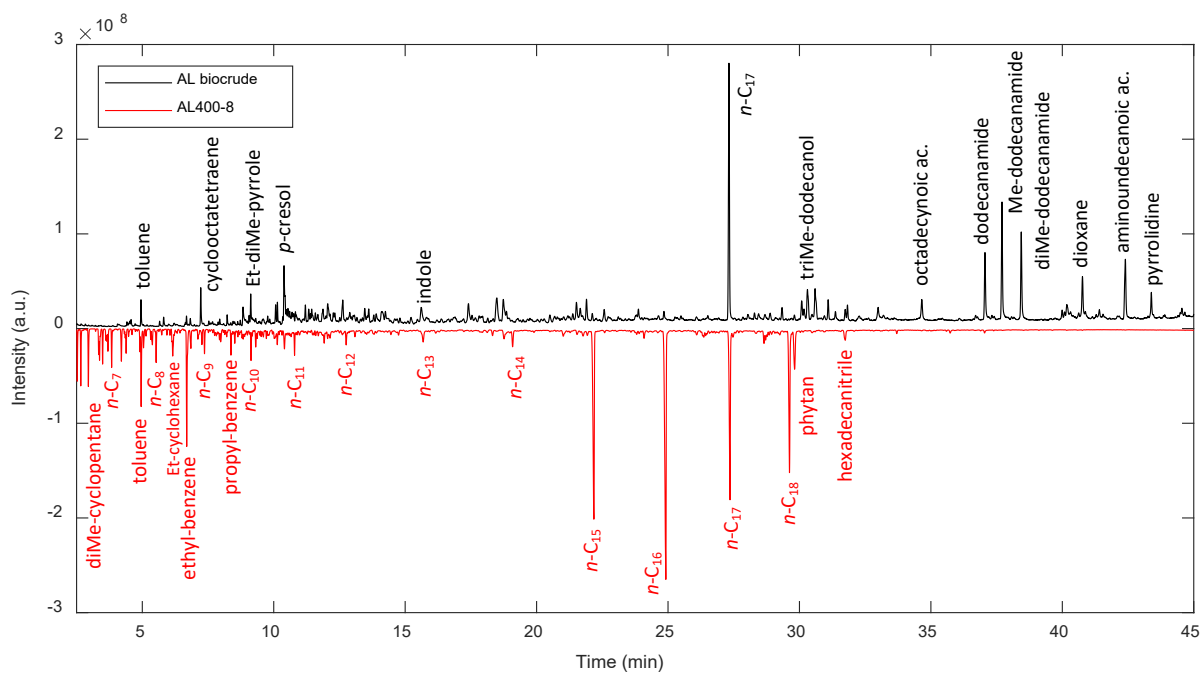

FIGURE 5

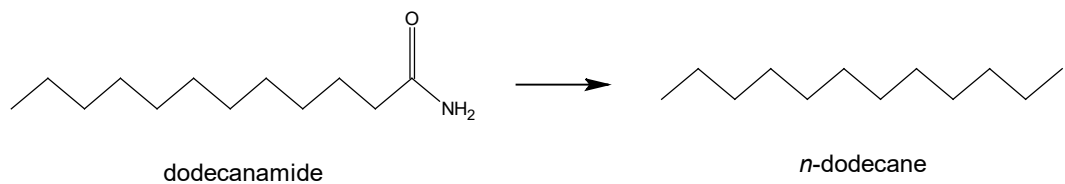

dodecanamide → n-dodecane

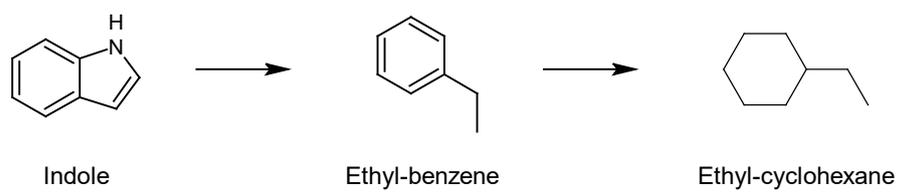

Indole → Ethyl-benzene → Ethyl-cyclohexane

FIGURE 6

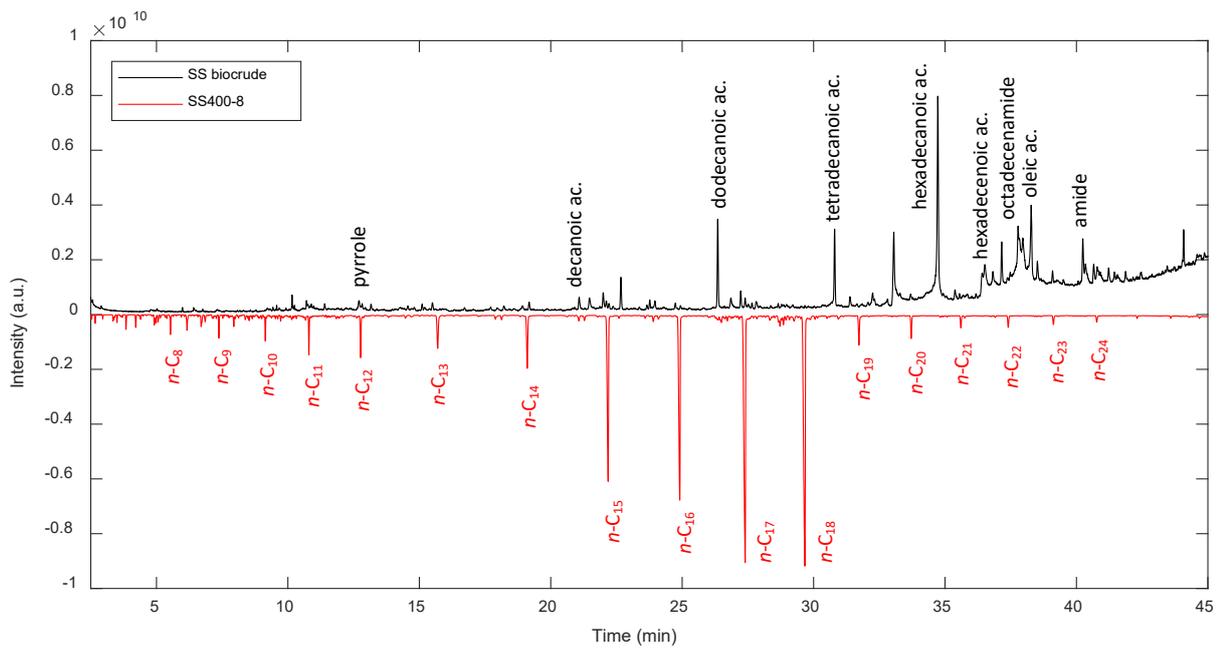

FIGURE 7

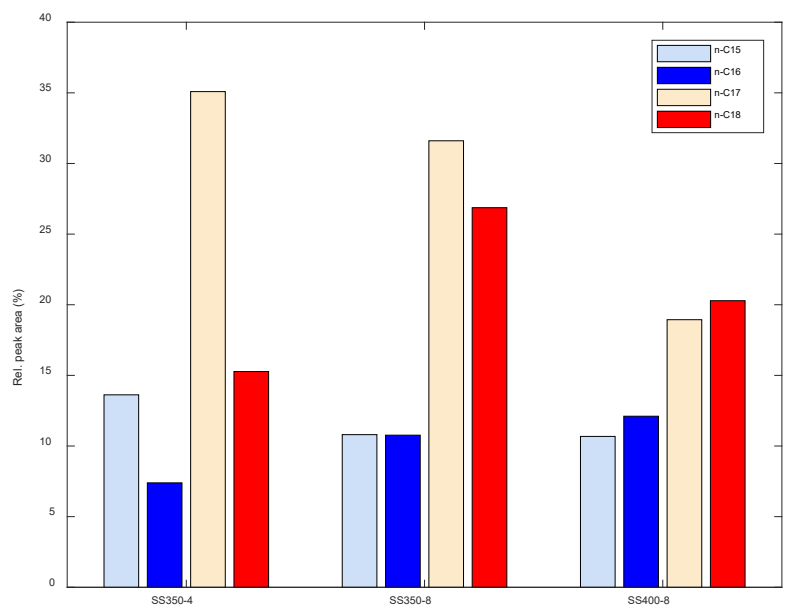

FIGURE 8

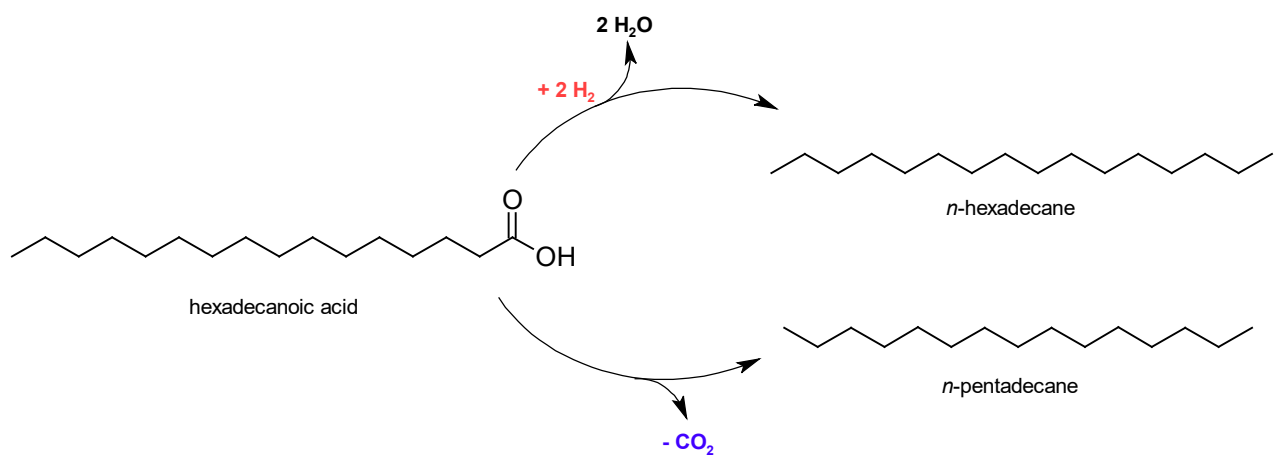

FIGURE 9

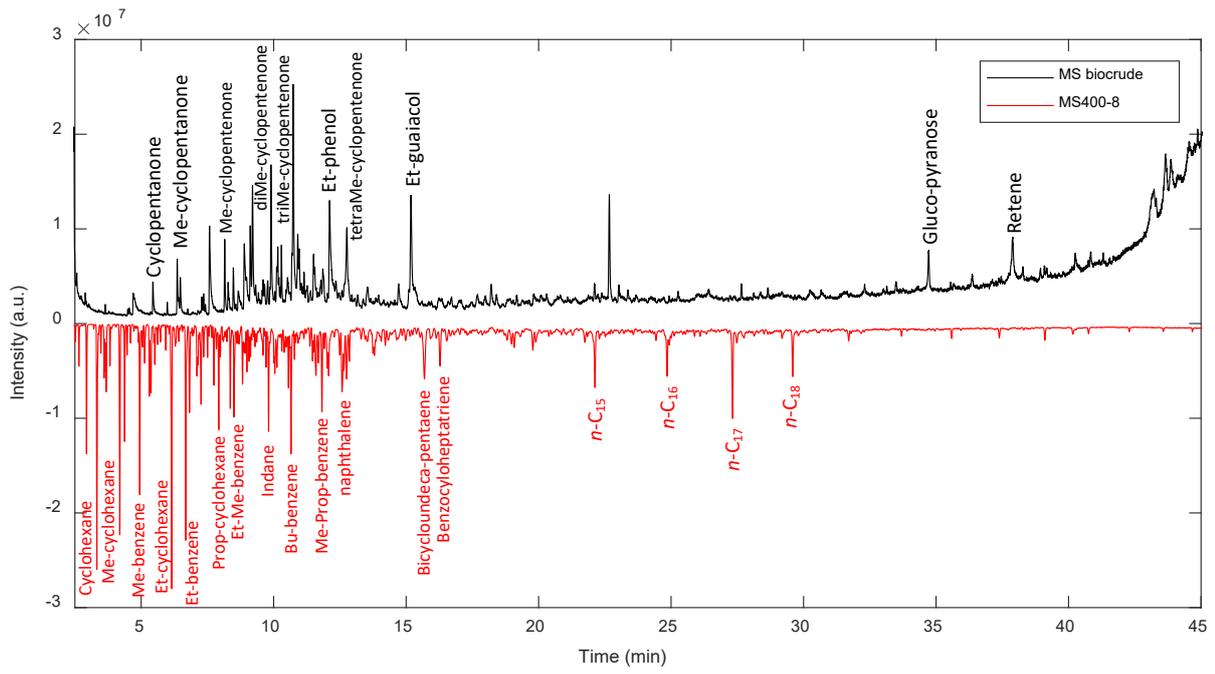